# Proton Dynamics in Protein Mass Spectrometry

*Jinyu Li[1#], Wenping Lyu[2,3,4#], Giulia Rossetti[2,5,6], Albert Konijnenberg[7], Antonino Natalello[8], Emiliano Ippoliti[2], Modesto Orozco[9,10], Frank Sobott[7,11,12], Rita Grandori[8], and Paolo Carloni[2,13]*</p>

[1]College of Chemistry, Fuzhou University, 350002 Fuzhou, China; [2]Computational Biomedicine, Institute for Advanced Simulation IAS-5 and Institute of Neuroscience and Medicine INM-9, Forschungszentrum Jülich, Jülich, Germany; [3]Faculty of Mathematics, Computer Science and Natural Sciences, RWTH-Aachen University, 52056 Aachen, Germany; [4]Cyprus Institute, Computat Based Sci & Technol Res Ctr, 2121 Aglantzia, Nicosia, Cyprus; [5]Department of Hematology, Oncology, Hemostaseology, and Stem Cell Transplantation, Faculty of Medicine, RWTH Aachen University, Aachen, Germany; [6]Jülich Supercomputing Centre (JSC), Forschungszentrum Jülich, D-52425 Jülich, Germany; [7]Biomolecular & Analytical Mass Spectrometry group, Department of Chemistry, University of Antwerp, Antwerpen, Belgium; [8]Department of Biotechnology and Biosciences, University of Milano-Bicocca, Piazza della Scienza 2, 20126 Milan, Italy; [9]Joint BSC-IRB Program on Computational Biology, Institute for Research in Biomedicine (IRB Barcelona), The Barcelona Institute of Science and Technology. Baldiri Reixac, 10, Barcelona 08028, Spain; [10]Departament de Bioquímica i Biomedicina. Facultat de Biologia. Universitat de Barcelona. Avgda Diagonal 647. Barcelona 08028. Spain; [11]Astbury Centre for Structural Molecular Biology, University of Leeds, Leeds LS2 9JT, United Kingdom; [12]School of Molecular and Cellular Biology, University of Leeds, LS2 9JT, United Kingdom; [13]JARA–HPC, 52425 Jülich, Germany.

[#]These authors contributed equally to this work.  *Corresponding author

# ABSTRACT

Native electrospray ionization/ion mobility-mass spectrometry (ESI/IM-MS) allows an accurate determination of low-resolution structural features of proteins. Yet, the presence of proton dynamics, observed already by us for DNA in the gas phase, and its impact on protein structural determinants, have not been investigated so far. Here, we address this issue by a multi-step simulation strategy on a pharmacologically relevant peptide, the N-terminal residues of amyloid-β peptide (Aβ(1-16)). Our calculations reproduce the experimental maximum charge state from ESI-MS and are also in fair agreement with collision cross section (CCS) data measured here by ESI/IM-MS. Although the main structural features are preserved, subtle conformational changes do take place in the first ~0.1 ms of dynamics. In addition, intramolecular proton dynamics processes occur on the ps-timescale in the gas phase as emerging from quantum mechanics/molecular mechanics (QM/MM) simulations at the B3LYP level of theory. We conclude that proton transfer phenomena do occur frequently during fly time in ESI-MS experiments (typically on the ms timescale). However, the structural changes associated with the process do not significantly affect the structural determinants.

**TOC**

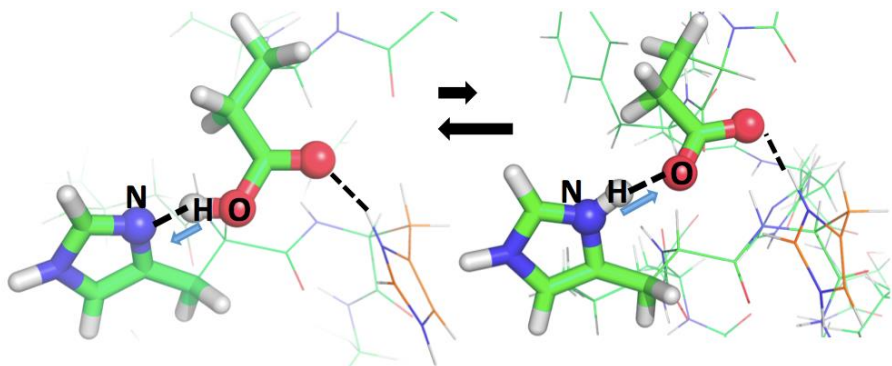

**MAIN TEXT**

Native electrospray ionization/ion mobility-mass spectrometry (ESI/IM-MS) is emerging as a powerful technique for capturing key structural features of proteins and their complexes.[1-7] Next to the mass-to-charge ratio (m/z), it provides the charge state distributions (CSD) and collision cross sections (CCS) for all species present in the gas phase. From these, one can extract their stoichiometry, topology, connectivity, dynamics and shape, as well as distribution of co-populated assembly and folding states.[8-16] In spite of lacking atomic resolution, ESI/IM-MS has distinct advantages over high-resolution methods such as X-ray crystallography and NMR spectroscopy as well as lower-resolution techniques such as cryo-electron microscopy (EM)[17] and tomography.[18] Indeed, it does not require crystallization and it is already sensitive at biomolecule concentrations well (roughly 1000 times) below those required for most of these techniques.[10, 19-20] In addition, it can characterize species distributions, i.e. co-populated folding and assembly states of proteins and complexes.

To further advance the impact of ESI/MS for structural biology, it is imperative to investigate the effect of charge in the absence of solvent, as experienced by the protein ions in the vacuum of the mass spectrometer, on protein structure and dynamics. In particular, proton dynamics between different ionizable residues (such as H, D, E, R, K) could play a role, as in nucleic acids,[21] which is so far unrecognized. Molecular simulations, performed by several groups including ours, may provide atomistic models of proteins under ESI-MS conditions,[22-30] consistent with available, low-resolution ESI/IM-MS structural data[29-30] and charge states.[22-23] Still, the key question on proton dynamics in protein – which requires a quantum mechanics treatment – remained elusive. To address this issue, we combine here a multi-step computational protocol already established for a variety of proteins in the gas phase[22-23] with a quantum mechanics/molecular mechanics (QM/MM) approach. Our multi-scale approach is applied to a peptide fragment containing the 16 N-terminal residues of the ~40 amino acid-long amyloid-β peptide (Aβ(1-16)), a promising therapeutic target to reduce cognitive deficits for Alzheimer's disease patients,[31] which is broadly studied by MS characterization in aspects of structure[32] as well as their aggregation behaviors.[33-34]

The maximum charge state (highest charge) of Aβ(1-16) is here predicted to be 4+ by our established hybrid Monte Carlo (MC)/MD-based protocol[22-23] (see Figure S1 in Supporting Information for details). Accordingly, the mass spectrum of the peptide shows a maximum charge state detectable at 4+ and a narrow CSD dominated by 3+ (main charge state, see Figure S2 in Supporting Information).

The lowest-energy protonation state, identified here by the MC/MD protocol at the main charge state ($q$=3+) in the gas phase, underwent MD simulations with three different initial microscopic conditions (called here "MD_gas1", "MD_gas2", "MD_gas3"). Simulation MD_gas1 was 0.129 ms-long, and equilibrium was reached after ~0.1 ms, as determined by Hess's cosine content analysis[35] (Table S1 in Supporting Information), and time-evolution plots of the backbone Root-Mean-Square-Deviation (RMSD, Figure 1A), CCS (Figure 1B), and end-to-end Cα distance (Figure 1C). Such a relaxation timescale compares well with experiments based on native electron capture dissociation.[36-37] Upon dehydration, the radius of gyration ($R_g$) and the solvent accessible surface area (SASA) decrease, with respect to the solution structure solved by NMR,[31] by 13% and 15%, respectively (Table S2 in Supporting Information). The SASA reduction indicates a moderate compaction of protein conformation, due to the absence of hydrophobic/hydrophilic interactions with interfacial water molecules in gas phase. At the same time, the number of intramolecular hydrogen bonds of Aβ(1-16) significantly increases by ~80% (Table S2 and MD1 in Table S3 in Supporting Information). These features have also been observed for other biomolecules.[23, 26] The secondary structure shares similarities to that in aqueous solution (Table S2 in Supporting Information).[31] In particular, the gas-phase structure preserves the turn centered at residues 7-8 [31] (Figure 1D). However, the N-terminal region is $3_{10}$-helical in water solution, as established by NMR measurements[31] while it is α-helical in the gas phase. This could be related either to the change from gas to solution and/or to known bias of AMBER force field-based MD, which disfavors the $3_{10}$-helix relative to the α-phase at helical.[38-42]

The calculated CCS value is 531 ± 15 Å$^2$ (Figure 1B). This value is in agreement with the experimental value measured in this work (539 ± 16 Å$^2$) for the same charge state (Figure S3 in Supporting Information). The structural properties from the other, slightly shorter (0.12 ms), MD simulations ("MD_gas2" and "MD_gas3") are overall similar to that of "MD_gas1" simulation, in spite of a significant variability of the secondary structure: the calculated values of CCS, SASA, HBs and $R_g$ differ from those calculated from the "MD_gas1" simulations by 8% or less (Table S2).

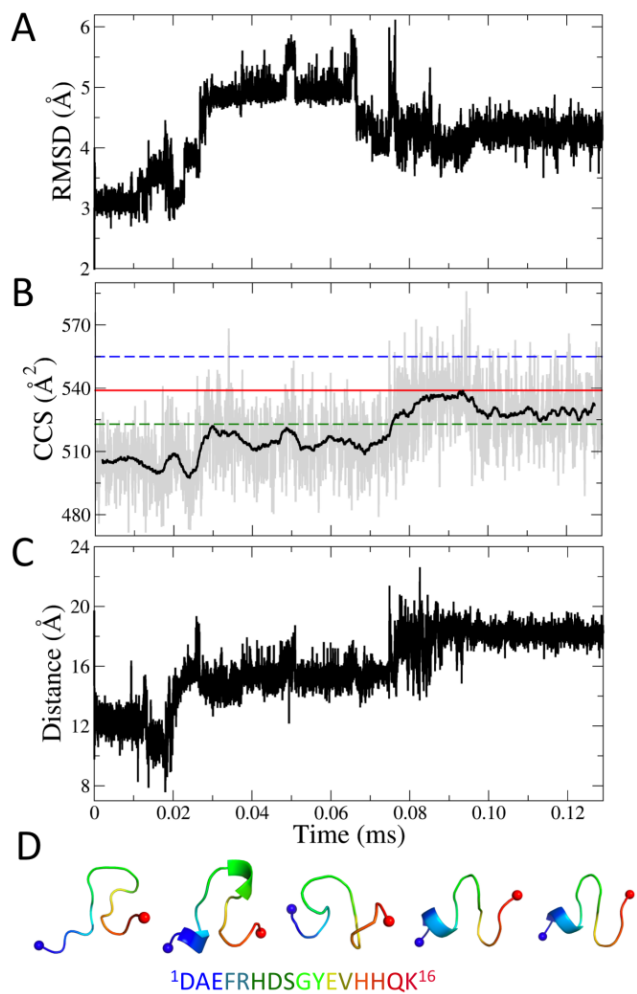

**Figure 1.** MD simulations in the gas phase of Aβ(1-16) at its main charge state (3+) found in ESI/(IM-)MS experiments. Result from the first simulation performed here (called "MD_gas1") are reported. (A)-(C), Time-dependence of: backbone atoms RMSD from the starting conformation (A); CCS values, where the experimental CCS at main charge state is indicated by a red solid line and its error bar is indicated by the dashed lines (B); distance between Cα atoms of N- and C-terminal residues (C). From left to right, conformations at 0 ms, 0.02 ms (RMSD=2.8 Å), 0.06 ms (4.9 Å), 0.09 ms (4.2 Å) and 0.12 ms (4.1 Å) (D). The N- and C-terminus are indicated by blue and red spheres, respectively.

We next switched to a QM description at the B3LYP level of theory of the ionizable residues of the protein forming intramolecular interactions (E3, H13, and H14) – E3 and H13 being neutral and H14 doubly protonated in our MD model (Figure 2A-B). The rest of the system is described, as before, using the AMBERff99SB-ILDN force field.[43-46] To capture the possible impact of the conformational differences on proton dynamics, three 12 ps-long QM/MM simulations were performed, starting from the representative conformations of the first three most populated clusters of the MD trajectory ("Traj_1", "Traj_2", "Traj_3", with populations of 88.1%, 5.7%, 2.3%, respectively). These clusters covered almost all (>96%) of the conformational space sampled in the MD (Figure S7).

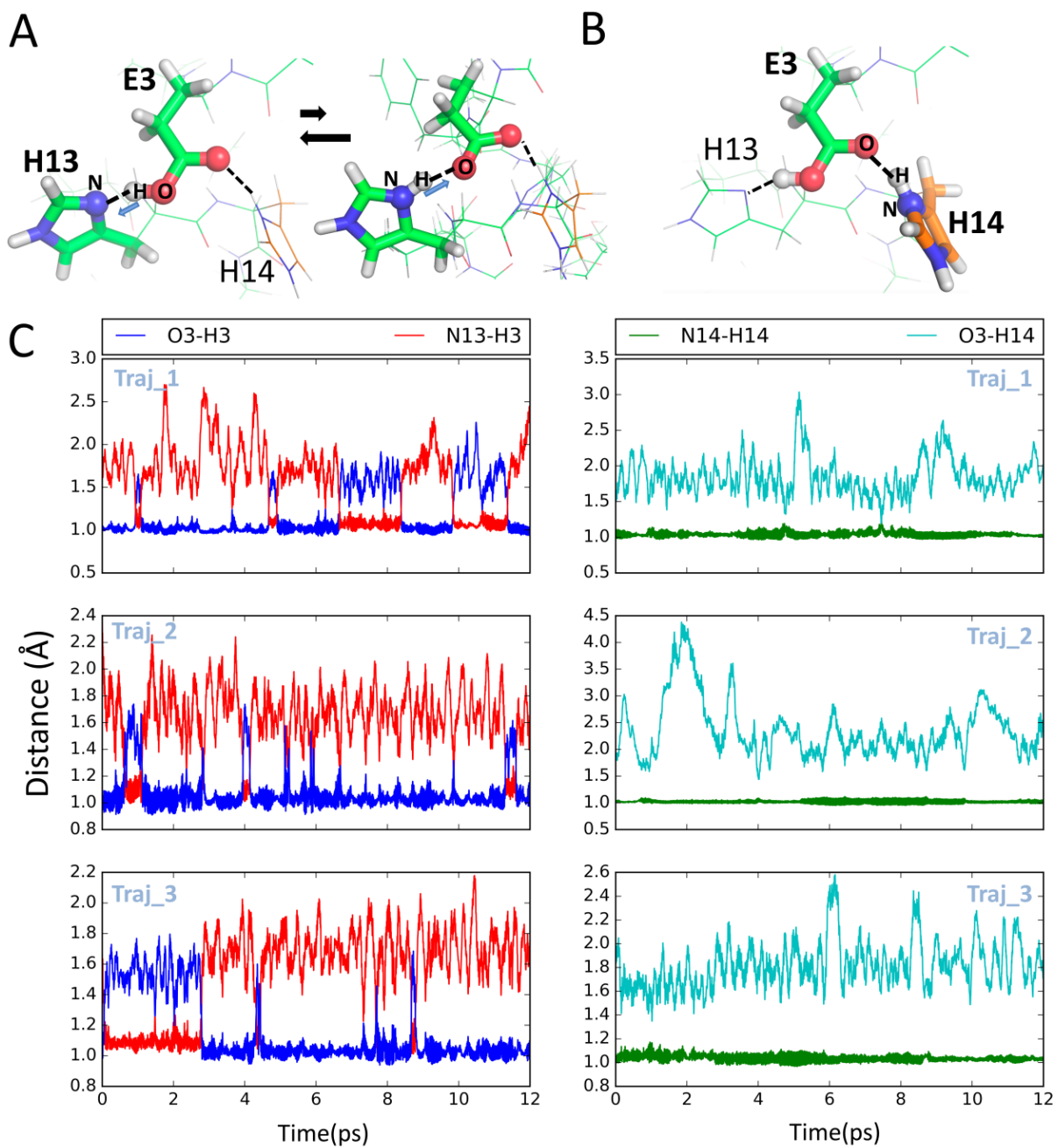

**Figure 2.** QM/MM simulations of the Aβ(1-16) peptide, starting from structures representative of the three most populated clusters of "MD_gas1" ("Traj_1", "Traj_2", "Traj_3", respectively). Conformations showing H13- (A), H14- (B) E3 hydrogen bonds (black dashed lines). (C) Proton atomic distance (in Å) from O3 and N13 (O3-H3 and N13-H3, respectively) and from N14 and O3 (N14-H14 and O3-H14, respectively) plotted as a function of time, for each QM/MM simulation.

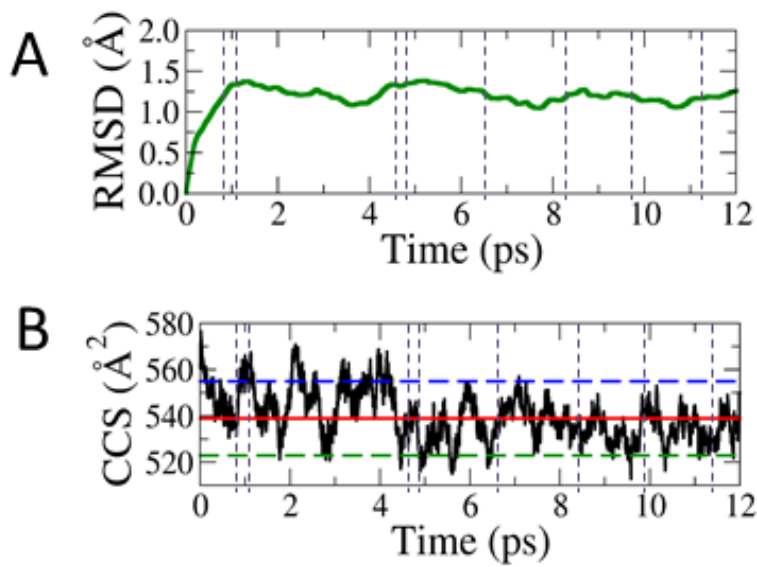

**Figure 3**. Time evolutions of the backbone atoms RMSD (A) and the CCS values (B) of Aβ(1-16) peptide in the QM/MM simulation starting from the representative conformation of the first cluster (the other simulations are reported in Fig. S8). The experimentally measured CCS at the main charge state (Fig. S3) is indicated by a red solid line. Its error bar is indicated by blue and green dashed lines. The occurrence of proton transfer is indicated by vertical dashed lines.

In all cases, we observed proton transfer between E3 and H13 more than half a dozen times (Figure 2C). No proton transfer occurred between E3 and H14. The most populated protonation state by far (79% overall in our three simulations) is the protonation state with both E3 and H13 being neutral (Table S4). This result is in line with previous experiments[47-52] and calculations,[53] showing that the ionic salt bridges are not particularly favored in the gas phase for the amino acids with weak or moderate proton affinities (such as histidine in this case). But the novel observation of an ionic salt bridge from the less populated protonation state (21%, Figure S6) leads us to suggest that the neutral and ionic forms of interactions do co-exist in the gas phase, interconverting in the ps time scale.

The proton transfer process does not affect the structure of the protein significantly: A weighted average of the three observed CCS resulting from the QM/MM simulation (544 ± 15 Å$^2$, 489 ± 8.2 Å$^2$, 482 ± 5.1 Å$^2$, for Traj_1-3, respectively) gives a value of 539 ± 14 Å$^2$ (Figure 3 and Figure S8). This turns out to be in agreement with the experimental ones, (539 ± 16 Å$^2$) (see Figure S3) and it is very similar to those observed in classical gas-phase MD, 531 ± 15 Å$^2$. The weighted average of the donor-acceptor distance of the hydrogen bond between E3 and H13 is 2.8 ± 0.2 Å (2.8 ± 0.2 Å, 2.7 ± 0.14 Å and 2.7 ± 0.1 Å, for "Traj_1-3", respectively), which is very similar to what observed in classical MD in the gas phase (2.9 ± 0.5 Å).

Combination of experiments, classical dynamics and for the first time dynamical quantum mechanical simulations provides a complete picture of the nuclei and electron dynamics of proteins in the gas phase during an electrospray experiments. First, our predictions of the maximum charge and of the CCS are fully consistent with experimental data measured here, validating our computational protocol. Second, our QM/MM simulations at the B3LYP level of theory for the QM part points to the importance of proton transfer, for the first time discussed here in proteins. We found that contrary to general assumptions in the field, electron dynamics is not negligible leading to unpredicted changes in the topology of the protein related to previously uncharacterized, easy and fast proton-transfer events. Our results strongly suggest that proteins in ESI/IM-MS experiments are not expected to behave as single covalent entities, with well-defined charge positions, but as an ensemble of charge sub-states generated by a flux of proton across ionizable residues.[1]

Several limitations associated with the QM/MM simulations should be discussed here. First, the QM model consists of residues E3, H13, and H14, without inclusion of residues interacting with them. Including those residues would from one hand increase the computational cost. On the other hand it is not expected to change the main result of the simulations, namely the presence of proton dynamics in the Aβ(1-16) peptide in ESI/IM-MS experiments. Next, one would like to use as accurate levels of theory as possible to describe the QM region. However, using high accurate methods, such as coupled cluster,[54] would make the calculations extremely expensive, most likely without changing the main findings here. Indeed, B3LYP, the level of theory used here, is known to perform reasonably well for H-bonding and proton transfer processes.[55] Finally, one could use polarizable force fields.[56] These could include the response of the protein frame electronic cloud to change of the QM electronic structure during proton dynamics. However, the reliability of such force fields for ESI-MS simulations, in contrast to the standard biomolecular ones,[22-23, 43-46, 57] still needs to be tested. Therefore, we have opted here for the AMBER ff99SB-ILDN force field,[43-46] successfully used by us in a variety of simulations of biomolecules in the gas phase.[22-23, 57-59] Therefore, we expect the main findings reported here to be confirmed by a computational effort even larger than the considerable one employed here.

---

[1]This is not the case for the scarcely populated protomer: our QM/MM simulations show that the peptide does not rearrange in this protonation state and it makes no sense to follow structural rearrangements of the system on a longer time-scale than that (ps) associated with proton dynamics.

In conclusion, our multi-scale molecular simulations reproduce the experimental maximum charge state by ESI/IM-MS. They also predict fairly well the CCS measured here. The predicted overall structure of the peptide upon dehydration differs subtly from the one calculated in solution. B3LYP functional-based QM/MM simulations uncover proton dynamics between E3 and H13 proton sites. This affects the charge of the residues involved in the process, while mostly preserving the structural determinants. In particular, the CCS calculated by QM/MM starting from the representatives of the most populated clusters is also in line with experimental data. Proton transfer processes as those uncovered here for Aβ(1-16), as well as for DNA,[21] in ESI/IM-MS conditions have never been observed in solution. Hence, we anticipate here that proton dynamics is a previously unrecognized fingerprint of biological structures during ESI/IM-MS experiments.

**MATERIALS AND METHODS**

**Experimental Material.** The peptide was synthetized according to the sequence DAEFRHDSGYEVHHQK-Ome by Proteogenix (Schiltigheim, France). The deconvoluted, average mass is 1968.768 (± 0.3311) Da (calculated average mass 1969.034 Da).

**Nano-ESI-MS.** Nano-ESI-MS analyses on 1mM Aβ(1-16) in 10 mM ammonium acetate pH 7.4 were performed under "native" conditions on a hybrid quadrupole-time-of-flight mass spectrometer (Qstar Elite; ABSciex, Framingham, MA) equipped with a nano-ESI ionization sample source. Metal-coated borosilicate capillaries (Proxeon, Odense, Denmark), with medium-length emitter tips of 1μm internal diameter, were used to infuse the samples. The instrument was calibrated by the standard Renin-inhibitor solution (ABSciex, Framingham, MA) on the intact molecular ion $(M+2H)^{2+}$ (879.97 Da) and its fragment $(F+H)^+$ (110.07 Da). Data were acquired in positive-ion mode with ion-spray voltage 1.2 kV and declustering potential 80 V, and were averaged over 2-min acquisitions. The interface was kept at room temperature (interface heater off).

**IM-MS.** IM-MS was performed on a Synapt G2 Q-TWIMS-TOF instrument (Waters, Manchester, U.K.) using $N_2$ as the drift gas. Ions were generated under "native" conditions by nano-ESI using in-house prepared gold-coated borosilicate glass needles. Critical voltages throughout the instrument were 1.2 kV capillary voltage, 25 V sampling cone, 0 V extraction cone, 4 V trap collision energy, 42 V trap DC bias and 0 V transfer collision energy. Pressures throughout the instrument were 2.82, 2.41 x $10^{-2}$, 3.07 and 2.53 x $10^{-2}$ mbar for the source region, trap collision cell, ion mobility cell, and transfer collision cell, respectively. All mass spectra were calibrated using 10 mg/ml CsI, and ion mobility drift times were calibrated against polyalanine clusters of known CCS to obtain experimental values.[60] The error in the CCS measurements was determined by repeat measurements under slightly different tuning conditions (variations in trap DC bias, wave velocity and wave height), which dictate the separation power as well as the duration of the experiment (see Figure S3). The variation between measurements while changing settings lies within 3%, whereas repeated experiments with similar settings varied by less than 0.3%.

**MD simulations of Aβ(1-16) peptide in aqueous solution.** We performed classical MD simulations in aqueous solution based on the twenty structures of the NMR conformational ensemble of Aβ(1-16) (Ac-$^1$DAEFRHDSGYEVHHQK$^{16}$-NH$_2$, PDB ID: 1ZE7[31]). The structure showing the best agreement with the averaged properties of the ensemble, out of all the twenty NMR structures present in the PDB (see Table S3), was selected as the initial structure for the MD simulations. The protonation states of residues in solution at neutral pH were assigned according to the corresponding p$K_a$ values calculated by using the H++ webserver.[61] As a result, the three histidine residues, H6, H13, and H14 were protonated at Nε nitrogen atoms; R5 and K16 were positively charged; and D1, E3, D7, and E11 were negatively charged. The peptide was inserted into a cuboid with each edge length of 52 Å containing 50 mM NaCl (corresponding to the salt concentration used in the NMR study of the peptide[31]) and ~4,400 water molecules. The overall system was neutral. The AMBER ff99SB-ILDN force field[43-46] and TIP3P force field[62] were used for the peptide and ions, and for water, respectively. Periodic boundary conditions were applied. Electrostatic interactions were calculated using the Particle Mesh-Ewald (PME) method,[63] and van der Waals and Coulomb interactions were truncated at 10 Å. All bond lengths were constrained using the LINCS algorithm.[64] First, the systems underwent 1000 steps of steepest-descent energy minimization with 1000 kJ·mol$^{-1}$·Å$^{-2}$ harmonic position restraints on the protein complexes, followed by 2500 steps of steepest-descent and 2500 steps of conjugate-gradient minimization without restraints. The systems were then gradually heated from 0 K up to 298 K in 20 steps of 2 ns. After that, four independent 1000 ns-long MD simulations were carried out in the canonical ensemble (298 K, 1 bar and 2 fs time-step) with different microscopic initial conditions (MD1 to MD4, hereafter). Constant temperature and pressure conditions were achieved by coupling the systems with a Nosé-Hoover thermostat[65-66] and an Andersen-Parrinello-Rahman barostat.[67] All the calculations of classical MD in this work were carried out using the GROMACS 4.5.5 code.[68] The CCS of the peptide was calculated using the trajectory method[69] implemented in the MOBCAL code.[70-71] Backbone atoms' RMSDs are reported in Figure S4.

**Determination of the lowest-energy protonation states of Aβ(1-16) peptide in the gas phase.** Our force field-based hybrid MC/MD protocol[22-23] was used for the determinations of the most probable protonation state of Aβ(1-16) in each charge state from $q=0$ to $q=4+$ in the gas phase. This procedure was implemented by considering that protons are mostly exchanged among a few sites, i.e. R, K, H, Q, E, and D side chains.[72] Hence, we protonated and deprotonated only these groups. We used the AMBER ff99SB-ILDN[43-46] force field augmented with a key modification that allows for proton exchange.[22-23] We previously showed that three different force fields (GROMOS41a1,[73] AMBER99,[74] and OPLS/AA[75]) give the same most probable protonation states for nine proteins of different size and fold, when the calculations were limited to protonation states containing the ionized residues common to all of the three force fields.[22] The modified force field was shown to successfully reproduce protonation-state energetics for folded peptides, proteins and a protein complex[22-23, 76] as calculated with density functional theory (DFT)-based simulations with dispersion corrections (for details see Ref. [23]).

We first optimized the temperature used in the MC/MD simulations. As the internal temperature of ions just emitted from droplets is associated with uncertainties, there is, up to now, no clear connection between simulation and experimental temperatures.[77] Specifically, 5 ns-long MD simulations in the gas phase were carried out on the Aβ(1-16) peptide for a randomly generated protonation state with temperatures of 300, 350, 400, 450, 500, 550, and 600 K coupling with a Nosé-Hoover thermostat.[65-66] By calculating the backbone atoms' RMSDs of the peptide obtained from these MD simulations at various temperatures (Figure S5A), we verified that only the structures for tested temperatures lower than 550 K are conserved. Therefore, 500 K was selected for the MC/MD simulations, since such temperature represents a good compromise between structural preservation and conformational sampling (Figure S5B). Our MC/MD approach is expected to discriminate between high- and low-energy protonation states, but not to capture small energy differences on the order of 10 kJ/mol.[22, 76] Hence, we considered the lowest-energy protonation state for each charge state ($q$=0 to $q$=4+), along with the protonation states whose energy differences from the lowest-energy one are less than 10 kJ/mol. This leads to the identification of one or two low-energy protonation states for each charge state of Aβ(1-16).

**MD simulations of Aβ(1-16) peptide in the gas phase.** We carried out three sets of independent MD simulations on the lowest-energy protonation state of the main charge state at 298 K in the gas phase (MD_gas1 to MD_gas3) by using the Nosé-Hoover thermostat.[65-66] The simulations covered 0.129 ms, 0.12 ms, and 0.12 ms, respectively. They differ for the initial velocities. The calculations were based on the AMBER ff99SB-ILDN force field.[43-46] AMBER ff99SB-ILDN is one of the most extensively used force fields for gas-phase simulations.[58-59] Moreover, compared with other force fields (e.g. CHARMM[78] and OPLS/AA[75]), the AMBER

ff99SB-based force field has reproduced best the secondary structure propensity of peptides with 3$_{10}$-helix (as the case of Aβ(1-16) studied here).[38]

**QM/MM simulations.** The three most representative structures obtained from the equilibrated trajectories of the gas-phase MD simulations were used as starting structures for B3LYP functional-based QM/MM simulations. Three ionizable residues form by far the most persistent H-bond interactions during the entire classical MD (E3, H13, and H14, see Table S6). Hence, these three residues are the most suitable groups to study proton dynamics. They were included in our QM region. In the QM part, the electronic wave function was expanded on a plane-wave basis set up to an energy cut-off of 90 Ry. This part was treated at DFT level and only the valence electrons were treated explicitly while the core ones were described through norm-conserving Troullier-Martins pseudopotentials.[79] The Kohn-Sham equations were solved using the B3LYP exchange-correlation functional.[80-81] Periodic boundary conditions were applied to the entire QM/MM box, while isolated system conditions in the QM part were imposed by using the Martyna-Tuckerman scheme for the Poisson solver.[82] Grimme's empirical corrections[83] were applied to provide an inexpensive yet reliable description of the van der Waals interactions. The dangling bonds in between the QM and MM regions[84] were saturated using an adapted monovalent carbon pseudopotential. The MM part includes the rest of the system and was described by the same force field as in the MD simulations. Constant temperature conditions were achieved by using the Nosé-Hoover chain thermostat with a reference temperature of 298 K. The electrostatic coupling between the QM and MM part were calculated using the fully Hamiltonian hierarchical approach of Ref.[85-86] In particular, the same QM/MM interface developed by that group was employed to couple the CPMD code (http://www.cpmd.org/,

Copyright IBM Corp 1990-2008, Copyright MPI für Festkörperforschung Stuttgart 1997-2001) with the classical MD engine of the GROMOS code.[87]

## ASSOCIATED CONTENT

**Supporting Information:** The Supporting Information is available free of charge on the ACS Publications website. Nano-ESI-MS and nano-ESI-IM-MS spectra, analysis of Hess's cosine content, analysis of MD simulations in solution and in the gas phase, the lowest-energy protonation states in the gas phase, prediction of the maximum charge state, determination of the temperature for MC/MD simulations, cartoon representation of the peptide structure obtained from QM/MM simulations after proton transfer (PDF).

## AUTHOR INFORMATION

**Corresponding Author**

*E-mail: p.carloni@fz-juelich.de

**Author Contributions**

JL and WL run the simulations and contributed equally to this work. AK and AN performed the experiments. All authors were contributed to the data analysis. The manuscript was written through contributions of all authors. All authors have given approval to the final version of the manuscript. The authors declare no competing financial interest.

## ACKNOWLEDGMENTS

WL is supported by the funding from the Horizon 2020 research and innovation program of the European Commission under the Marie Sklodowska-Curie grant agreement No. 642069. JL is supported by a grant from the National Science Foundation of China No. 21603033. This work makes use of results or expertise provided by BioExcel CoE (www.bioexcel.eu), a project funded by the European Union contract H2020-EINFRA-2015-1-675728. The authors gratefully acknowledge the computing time granted on the supercomputer JURECA[88] at Jülich Supercomputing Centre (JSC). The authors would like to thank the Hercules foundation for funding of the Synapt G2 instrument. PC acknowledges funding from the Human Brain Project.

# REFERENCES

1. Sharon, M. Biochemistry. Structural MS pulls its weight. *Science* **2013,** *340* (6136), 1059-60.
2. Kocher, T.; Superti-Furga, G. Mass spectrometry-based functional proteomics: from molecular machines to protein networks. *Nature methods* **2007,** *4* (10), 807-15.
3. Taverner, T.; Hernandez, H.; Sharon, M.; Ruotolo, B. T.; Matak-Vinkovic, D.; Devos, D.; Russell, R. B.; Robinson, C. V. Subunit architecture of intact protein complexes from mass spectrometry and homology modeling. *Accounts of chemical research* **2008,** *41* (5), 617-27.
4. Hopper, J. T.; Robinson, C. V. Mass spectrometry quantifies protein interactions--from molecular chaperones to membrane porins. *Angewandte Chemie* **2014,** *53* (51), 14002-15.
5. Zhong, Y.; Han, L.; Ruotolo, B. T. Collisional and Coulombic unfolding of gas-phase proteins: high correlation to their domain structures in solution. *Angewandte Chemie* **2014,** *53* (35), 9209-12.
6. Ma, X.; Lai, L. B.; Lai, S. M.; Tanimoto, A.; Foster, M. P.; Wysocki, V. H.; Gopalan, V. Uncovering the stoichiometry of Pyrococcus furiosus RNase P, a multi-subunit catalytic ribonucleoprotein complex, by surface-induced dissociation and ion mobility mass spectrometry. *Angewandte Chemie* **2014,** *53* (43), 11483-7.
7. Konijnenberg, A.; Butterer, A.; Sobott, F. Native ion mobility-mass spectrometry and related methods in structural biology. *Biochim Biophys Acta* **2013,** *1834* (6), 1239-56.
8. Konijnenberg, A.; Ranica, S.; Narkiewicz, J.; Legname, G.; Grandori, R.; Sobott, F.; Natalello, A. Opposite Structural Effects of Epigallocatechin-3-gallate and Dopamine Binding to alpha-Synuclein. *Anal Chem* **2016,** *88* (17), 8468-75.
9. Konijnenberg, A.; Yilmaz, D.; Ingolfsson, H. I.; Dimitrova, A.; Marrink, S. J.; Li, Z.; Venien-Bryan, C.; Sobott, F.; Kocer, A. Global structural changes of an ion channel during its gating are followed by ion mobility mass spectrometry. *Proc Natl Acad Sci U S A* **2014,** *111* (48), 17170-5.
10. Hyung, S. J.; Ruotolo, B. T. Integrating mass spectrometry of intact protein complexes into structural proteomics. *Proteomics* **2012,** *12* (10), 1547-1564.
11. Sharon, M. How Far Can We Go with Structural Mass Spectrometry of Protein Complexes? *J. Am. Soc. Mass Spectrom.* **2010,** *21*, 487-500.
12. Pukala, T. L.; Ruotolo, B. T.; Zhou, M.; Politis, A.; Stefanescu, R.; Leary, J. A.; Robinson, C. V. Subunit Architecture of Multiprotein Assemblies Determined Using Restraints from Gas-Phase Measurements. *Structure* **2009,** *17*, 1235-1243.
13. Jurneczko, E.; Barran, P. E. How useful is ion mobility mass spectrometry for structural biology? The relationship between protein crystal structures and their collision cross sections in the gas phase. *The Analyst* **2011,** *136* (1), 20-8.
14. Zhou, M.; Robinson, C. V. When proteomics meets structural biology. *Trends in biochemical sciences* **2010,** *35* (9), 522-9.
15. Schmidt, C.; Robinson, C. V. A comparative cross-linking strategy to probe conformational changes in protein complexes. *Nature protocols* **2014,** *9* (9), 2224-36.
16. Politis, A.; Stengel, F.; Hall, Z.; Hernandez, H.; Leitner, A.; Walzthoeni, T.; Robinson, C. V.; Aebersold, R. A mass spectrometry-based hybrid method for structural modeling of protein complexes. *Nat Methods* **2014,** *11* (4), 403-6.
22